\documentclass[12pt]{article}

\usepackage{amsmath,amssymb}

\usepackage{epsfig}

\begin{document}

\title{Fundamental cosmic strings}

\author{A.-C.~Davis$^1$ and T.W.B.~Kibble$^2$\\ \small
$^1$DAMTP, University of Cambridge, CMS, Wilberforce Road,
Cambridge CB3 0WA\\ \small
$^2$Blackett Laboratory, Imperial College, London SW7 2AZ}

\date{\small\today}

\maketitle

Cosmic strings are linear concentrations of energy that may be
formed at phase transitions in the very early universe.  At
one time they were thought to provide a possible origin for
the density inhomogeneities from which galaxies eventually
develop, though this idea has been ruled out, primarily by
observations of the cosmic microwave background (CMB). 
Fundamental strings are the supposed building blocks of all
matter in superstring theory or its modern version, M-theory. 
These two concepts were originally very far apart, but recent
developments have brought them closer.  The `brane-world'
scenario in particular suggests the existence of macroscopic
fundamental strings that could well play a role very similar
to that of cosmic strings.

In this paper, we outline these new developments, and also
analyze recent observational evidence, and prospects for the
future.

\section{Spontaneous symmetry breaking}

This is a feature of many physical systems, often accompanying
a phase transition.  For example, there are no preferred
directions in a liquid such as water --- it has complete
rotational symmetry.  But when it freezes the resulting
crystal \emph{does} have a definite orientation --- the full
rotational symmetry is broken, leaving only a very restricted
group of symmetries of the crystal.  It would be difficult for
a microscopic inhabitant of the crystal to infer the existence
of the larger symmetry!

When a liquid is cooled through its freezing point, crystals
of solid will start to form, often nucleated by small
impurities, but one cannot predict what their orientation will
be.  The choice is random, and different choices may be made
at different centres, so that when the entire liquid has
frozen, there may be mismatches, leading to grain boundaries
or more subtle defects where the crystal lattice is deformed. 
For instance, one can have an extra layer of atoms ending on a
linear `edge defect', where the crystal is strained.  Similar
`topological defects' occur in many theories of fundamental
particle interactions.

Spontaneous symmetry breaking is a ubiquitous feature of our
theories of fundamental particle interactions.  (See, for
example, \cite{Wei96}.)  The famous electroweak theory, for
which Sheldon Glashow, Abdus Salam and Steven Weinberg won the
1979 Nobel Prize, exhibits an underlying symmetry between the
carriers of the electromagnetic force and the weak force ---
the photon and the W and Z particles.  When we only had access
to low-energy experiments this symmetry was completely hidden,
its existence as difficult to guess as would be the full
symmetry of the crystal to its microscopic inhabitant.  The
symmetry \emph{is} apparent in very high-energy experiments, at
energy scales well above 100 GeV ($10^{11}$ electron volts),
but we live in a low-temperature phase where it is
spontaneously broken by the so-called Higgs mechanism which
imparts masses to the W and Z bosons, while leaving the photon
massless.

Following the success of the electroweak model, physicists
started to ask if the strong interactions too could be brought
within a single unified framework.  There is some experimental
support for this idea of `grand unification'.  The strengths
of the fundamental interactions are determined by three
`coupling constants' $g_1,g_2,g_3$, which, despite their name,
depend weakly (logarithmically) on energy.  If one
extrapolates from the low energies where they are measured one
finds that all three come together at an energy scale of about
$10^{16}$ GeV, strongly suggesting that something interesting
happens at that scale.  Several different `grand unified
theories' (GUTs) have been proposed to embody this idea, the most 
succesful involve a symmetry between bosons and fermions,
called supersymmetry.

Unfortunately particle energies of $10^{16}$ GeV are far
beyond any scale accessible to present or future laboratory
experiments, so GUTs are hardly likely to get the kind of
solid experimental support the electroweak theory received. 
There is however one place --- or rather time --- where we
believe such energies did occur, the very early universe in
the first fraction of a second after the Big Bang.  We know
that the temperature of the early universe, back to when it
was a few minutes old (the time of primordial nucleosynthesis),
was falling like the inverse square root of its age, $T\propto
1/\sqrt{t}$.  If we trace its history back still further, our
best guess is that its temperature would have been above the
electroweak phase transition, at a temperature of around
$10^{15}$ K, when it was less than a microsecond old.  Before
that the full electroweak symmetry would have been evident. 
Even further back, if the grand unification idea is correct,
it would have gone through a GUT transition, at the
unimaginably early age of $10^{-35}$ s.

Of course no observers were around to check these
speculations.  We have to rely instead on very indirect
evidence of these early phase transitions.  One clue could
come from looking for the characteristic topological defects
that might have formed in the process of spontaneous symmetry
breaking.  Such defects are often stable, so some at least of
them could have survived to much later times, perhaps even
to the present day.  Many types of defects are possible,
depending on the nature of the symmetry breaking --- point
defects (monopoles), linear defects (cosmic strings), planar
defects (domain walls), as well as combinations of these. 
However, the most interesting have turned out to be cosmic
strings.  (Reviews can be found in \cite{HinK95,VilS94,Raj03}.)

\section{Cosmic strings}

The simplest model that shows what cosmic strings might be
like is one with a single complex scalar field $\phi$, which
we can also think of as a pair of real fields $\phi_{1,2}$,
with $\phi=\phi_1+i\phi_2$.  The symmetry in this case is a
phase symmetry.  We assume that the Hamiltonian which defines
the dynamics of the field is invariant under the phase
rotation $\phi\to\phi e^{i\alpha}$, or equivalently
 \begin{equation}
 \begin{split}
&\phi_1\to\phi_1\cos\alpha-\phi_2\sin\alpha,\\
&\phi_2\to\phi_1\sin\alpha+\phi_2\cos\alpha.
 \end{split} 
 \end{equation}
In particular, there is a potential energy term $V$ which is a
function only of $|\phi|$, usually taken to be the `sombrero
potential'
 \begin{equation}
V=\frac{1}{2}\lambda(|\phi|^2-\eta^2)^2=
\frac{1}{2}\lambda(\phi_1^2+\phi_2^2-\eta^2)^2,
 \end{equation}
 \begin{figure}[htb]
 \centerline{\psfig{file=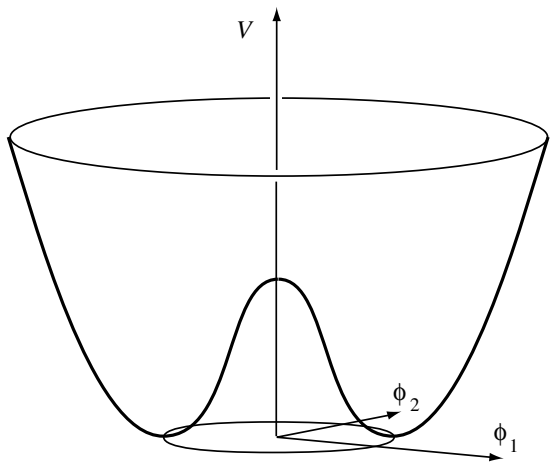}}
 \vspace*{8pt}
 \caption{The sombrero potential.}
 \end{figure}
where $\eta$ is a constant (see Fig.~1).  The important thing
to notice is that the minimum is not at $\phi=0$ but around the
circle $|\phi|=\eta$.  There is a degenerate ground state: any
of the points $\phi=\eta e^{i\alpha}$ around the circle
defines a ground state.

At high temperatures, there are large fluctuations in $\phi$
and the central hump in the potential is unimportant.  Then
there is obvious symmetry: fluctuations of $\phi$ in any
direction are equally likely.  But as the temperature falls
there comes a point where the energy is too low to permit
fluctuations over the hump.  Then the field tends to settle
towards one of the ground states.  Which point on the circle
of minima is chosen is a matter of random choice, like the
choice of direction of fall for a pencil balanced on its tip. 
The spontaneous choice then breaks the original symmetry.

When a large system goes through a phase transition like this,
each part of it has to make this random choice of the phase angle
$\alpha$, but as in the process of crystallization the choice
may not be the same everywhere --- one part of the system does
not `know' of the choice made in a distant part \cite{Kib76}. 
Because there are terms in the energy involving the gradient of
$\phi$, the phase angle will tend to become more uniform as
the system cools.  But this process may be frustrated; the
random choices may not fit together neatly.  In particular,
there may be linear defects --- cosmic strings (see Fig.~2) ---
around which the phase angle varies by $2\pi$ (or a multiple of
$2\pi$).
 \begin{figure}[htb]
 \centerline{\psfig{file=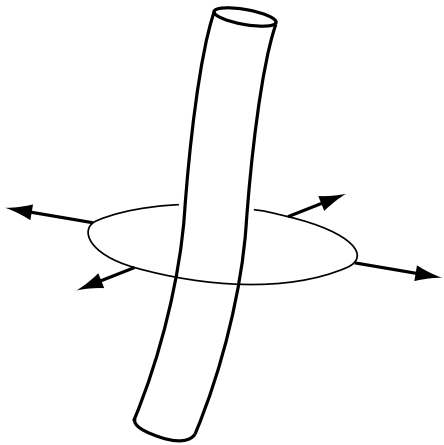}}
 \vspace*{8pt}
 \caption{A cosmic string.  The directions of the arrows
indicate the values of $\alpha$.  The field $\phi$ vanishes in
the core of the string.}
 \end{figure}

\section{Strings in the early universe}

The important thing about cosmic strings is their stability. 
Continuity of $\phi$ means that a string cannot simply come to
an end; it must form a closed loop or extend to infinity (or
at least beyond the region we can see).  For this reason,
strings, once formed, are hard to eliminate.  In the core of
the string, $\phi$ vanishes, so there is trapped potential
energy (as well as gradient energy).  So strings represent
trapped energy.  In fact, the core of the string may be
regarded as a relic of the earlier high-temperature phase, and
the energy density there is similar to what it was before the
transition.  To lower this energy, strings will tend to
shrink, smoothing out kinks, though at the same time they are
being stretched by the expansion of the universe.  They may
cross and exchange partners (see Fig.~3a) --- a process known
as `intercommuting', which creates new kinks.  A string may
also cross itself, forming a closed loop (see Fig.~3b).  Such
 \begin{figure}[htb]
 \centerline{\psfig{file=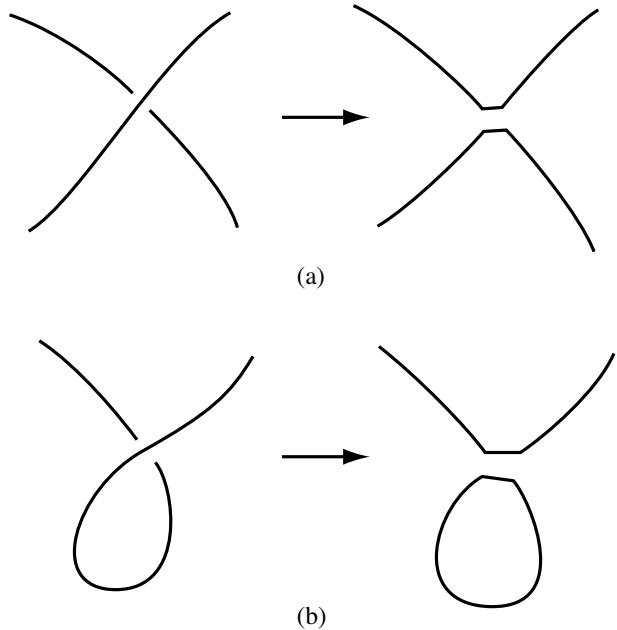}}
 \vspace*{8pt}
 \caption{a. Intercommuting of strings.  b. Formation of a
closed loop.}
 \end{figure}
loops may shrink and eventually disappear, but a long string
cannot do so directly.  If a random tangle of strings was
formed in the early universe, there would always be some
longer than the radius of the visible universe, so a few would
remain even today.

Analogous defects are formed in many condensed-matter systems
undergoing low-temperature phase transitions.  Examples are
vortex lines in superfluid helium, magnetic flux tubes in some
superconductors, and disclination lines in liquid crystals.  A
nematic liquid crystal, for example, consists of rod-shaped
molecules that like to line up parallel to each other. 
Everywhere in the liquid there is a preferred orientation, but
note that diametrically opposite directions are equivalent. 
Around a disclination line, the preferred orientation rotates
by $180^\circ$ (see Fig.~4).  Along the line, molecules do
not know which way to turn, and there is excess trapped
energy.  It is easy to see in this case too that a
disclination line cannot simply end.
 \begin{figure}[htb]
 \centerline{\psfig{file=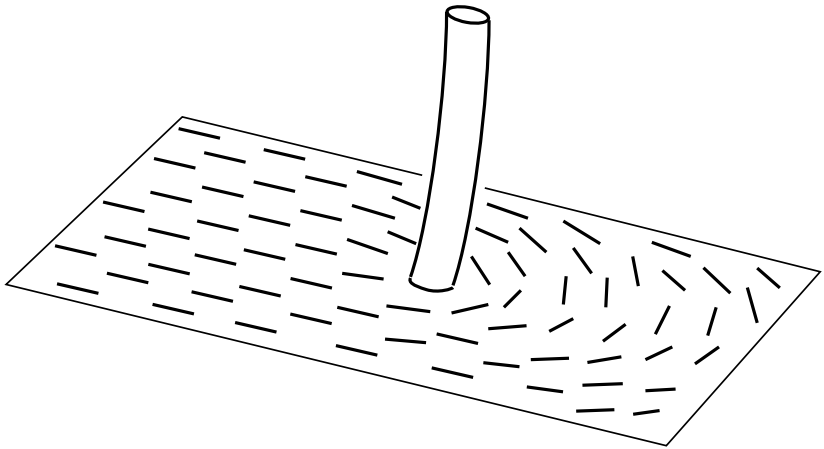}}
 \vspace*{8pt}
 \caption{A disclination line in a nematic liquid crystal.}
 \end{figure}

Experiments on cosmic strings would be impossible, even if we
knew for sure that they existed.  But because there are these
various analogues of cosmic strings, many interesting
experiments have been done testing various aspects of the
cosmic string formation and evolution scenario \cite{Kib02},
though of course none of this can tell us whether there really
are cosmic strings in the universe.  For that we have to turn
to astronomical observations, to which we shall return later.

In the late 80s and early 90s, cosmic strings generated a lot
of excitement among cosmologists because they seemed to offer
a plausible explanation for the origin of the density
inhomogeneities from which galaxies later developed.  Because
they represent a lot of trapped energy, cosmic strings
thrashing about in the early universe would significantly
perturb the matter distribution, and it is not hard to get at
least a rough estimate of how big the effect would be.  The key
parameter is the energy per unit length of the string which,
for reasons of relativistic invariance (the characteristic
speed of waves on the string is the speed of light, $c$), is
equal to the string tension $\mu$.  The strings are
exceedingly thin, but very massive.  Typically, for strings
produced at a possible GUT transition, the mass per unit
length $\mu/c^2$ would be of order $10^{21}$ kg m$^{-1}$; a
string of length equal to the solar diameter would be about as
massive as the Sun itself.  The gravitational effects
of a string are governed by a dimensionless parameter,
$G\mu/c^4$, where $G$ is Newton's constant.  In particular,
strings in the early universe would create density
perturbations in which the fractional change in density is
 \begin{equation}
\frac{\delta\rho}{\rho}\sim\frac{G\mu}{c^4}.
 \end{equation}
It happens that for GUT-scale strings, this ratio would be
roughly $10^{-6}$ or $10^{-7}$.  This is just the right order
of magnitude to seed galaxy formation.  In what follows, we
shall choose units in which $c=1$ and talk of the parameter
$G\mu$.

Unfortunately this nice idea, like so many others, succumbed
to the harsh realities of observation, in particular the
observations of the small anisotropies in the cosmic microwave
background.  Measurements made by the COBE (COsmic Background
Explorer) satellite and more recently by WMAP (the Wilkinson
Microwave Anisotropy Probe) have yielded very precise
information about these anisotropies.  In particular, the
angular power spectrum shows a series of peaks, so-called
`acoustic peaks', representing particular scales on which the
anisotropy is large (see Fig.~5).  The cosmic string scenario
 \begin{figure}[htb]
 \centerline{\psfig{file=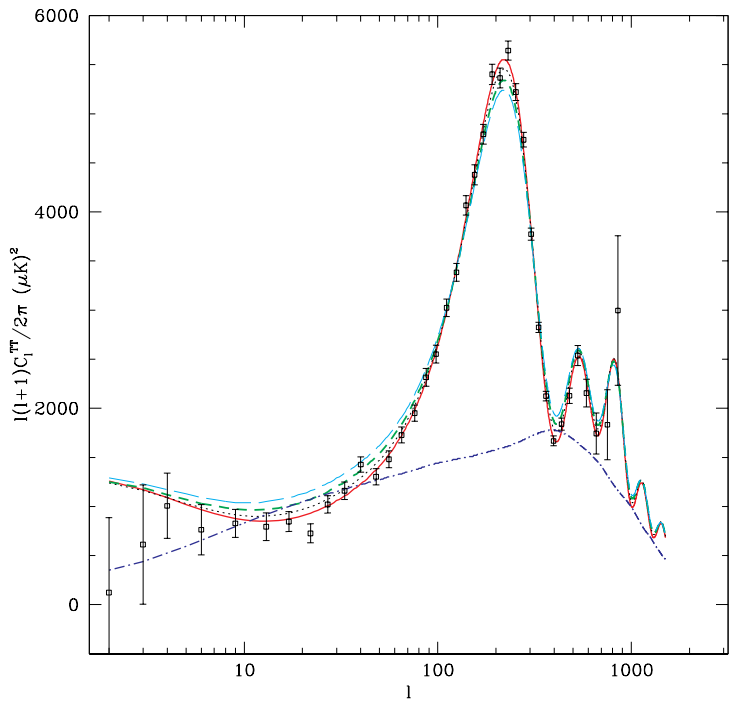}}
 \vspace*{8pt}
 \caption{Angular power spectrum of CMB \cite{Pog+03}.  The
solid (red) line corresponds to $B=0$, the dotted (black) line
$B=0.05$, the short-dash (green) line $B=0.1$, the long-dash 
(light blue) line $B=0.15$ and the dot-dash (dark blue) line to
$B=1$, where $B$ is the fraction of the power due to defects.}
 \end{figure}
has no explanation for these features, predicting instead a
single broad, flattish hump.  On the other hand, the peaks are
exactly what was predicted by the rival theory of inflation,
in which the origin of the density perturbations can be traced
to quantum fluctuations during an early period of accelerated
expansion.

So inflation won, and cosmic strings were relegated at best to
a minor supporting role, responsible for no more than 10\% of
the density perturbation at most.  Many people lost interest
in the idea.

\section{Superstring theory}

There has, however, been a revival of interest, stemming
largely from developments in our understanding of a very
different kind of string --- the fundamental strings of
superstring theory, or its more modern incarnation, M-theory. 
(For a recent review, see \cite{Gre00}.)

Fundamental string theory also originated in a search for
unification, in particular the unification of gravity with the
other interactions.  This has long been the holy grail of
theoretical physics, but has proved remarkably elusive.  A
major obstacle to creating a quantum theory of gravity
(unified or not) has been the appearance of infinities. 
\emph{All} quantum field theories are plagued by infinities,
but for the other interactions we have learned how to tame
them, by the process of `renormalization'.  This allows us to
extract meaningful, finite answers to physically significant
questions, hiding all the infinities in supposedly
unmeasurable `renormalization constants'.  This has never been
a wholly satisfactory procedure, but in any case it fails
entirely in the case of gravity.  No quantum version of
Einstein's theory of general relativity is renormalizable;
there are infinities that cannot be swept under the carpet.

The basic reason for the appearance of infinities is the fact
that we are dealing with point particles.  They appear even in
classical electromagnetic theory, for example, in the
`self-energy' of a charged particle: the potential energy
stored in a spherical distribution of charge goes to infinity
as its radius tends to zero.

This observation led to a very intriguing proposal: perhaps the
fundamental objects of our theory should not be point
particles, but extended objects, in particular strings.  The
basic idea is that all the particles we commonly think of as
elementary --- electrons, quarks, photons, and the rest ---
can be regarded as different modes or oscillation states of
a fundamental string.

Even with strings as the basic building blocks, it proved
difficult to eliminate all the infinities.  One feature that
made it easier was to incorporate \emph{supersymmetry} into
the theory.  This is a remarkable symmetry that relates bosons
(particles of integer spin in units of $\hbar$) to fermions
(with half-integral spin).  In a perfectly supersymmetric
world, every boson would be partnered by a corresponding
fermion of equal mass, and \emph{vice versa}.  Such partners
have never been seen, so if our world is fundamentally
supersymmetric, this symmetry too must be broken.  But the
great virtue of supersymmetry is that it removes a lot of
infinities.  This is because bosons and fermions often make
equal and opposite contributions, so that if they are exact
partners the infinities cancel.

One other strange feature was needed.  Superstring models
were eventually constructed that did seem to be free of all
infinities, but \emph{not} in the familiar four dimensions
(three of space and one of time).  The models were only
consistent in 10 dimensions (nine space and one time) --- or
even sometimes 26!  So the suggestion emerged that our
universe is fundamentally ten-dimensional, but six of the
dimensions are curled up very small, so that from our
macroscopic perspective it looks four-dimensional --- just as a
drinking straw looks one-dimensional on a large scale.

These fundamental strings, as originally envisaged, were very
different in many ways from cosmic strings.  Firstly, their
characteristic energy scale was much larger.  Gravity is
strongly energy-dependent, but would become as strong as the
other interactions at the so-called \emph{Planck scale},
corresponding to an energy around $10^{19}$ GeV, at least a
thousand times higher than the GUT scale.  Thus the parameter
$G\mu$ for a fundamental string was of order one. 
Moreover, the fundamental strings could never have extended to
macroscopic size: if you try to expand a fundamental string it
will simply break into several small pieces.

But this picture has changed in several ways.  There are other
mechanisms for reducing the dimension from 10 to 4, whose
effects are rather different, in particular the
\emph{brane-world} idea.  Strings are not the only localized
objects in a superstring theory.  There can be two-dimensional
\emph{membranes} or their higher-dimensional analogues, which
have come to be known as `$p$-branes' (of dimension $p$) --- a
particle is a 0-brane, a string a 1-brane, and so on.  We have
D-branes (or D$p$-branes), where the D denotes Dirichlet
boundary conditions (see \cite{Gaunt98} for a review). Essentially this 
means that in addition
to closed loops of fundamental string there may be open strings
whose ends are tied to D-branes. There are also anti-branes, 
$\overline{\mathrm{D}}$-branes.  A D$p$ brane and a
$\overline{\mathrm{D}p}$ have equal and opposite conserved
charges, which means that they attract each other. Open strings
usually give rise to matter fields while gravity comes from closed
loops. This means that matter may be trapped on a
D$p$-brane while gravity feels all the extra dimensions.

The brane-world concept has also emerged recently from M-theory.
M-theory is a conjectured umbrella theory that, in certain
limits, reduces to the five known string theories or to
supergravity. Supergravity was another attempt to unify all
the forces, including gravity and incorporating
supersymmetry and was formulated in $11$ dimensions! It was
shown that string theory at low energy is described by
eleven-dimensional supergravity with the eleventh dimension
compactified on an interval with $\mathbb{Z}_2$ (mirror)
symmetry. The two boundaries of space--time are
10-dimensional planes, on which  matter is confined. The
extra six dimensions are compactified, but these compact
dimensions are substantially smaller than the space  between
the two boundary branes.  Thus, viewed on a larger scale,
space--time looks five-dimensional with four-dimensional
boundary branes. This is an example of a brane world model
(see \cite{braxbd04} for a recent review).

More generally speaking, in brane-world models normal matter
is confined on a hypersurface (called a brane) embedded in a
higher-dimensional space (called the bulk).  Our universe may 
be such a brane-like object.  Within the brane-world scenario,
constraints on the size of extra dimensions become weaker
because the standard particles propagate only in three spatial
dimensions, while gravity (and some other exotic matter) 
can propagate in the bulk. Newton's law of gravity, however, is
sensitive to the presence of extra dimensions, and deviations
are expected on scales below the brane separation. Gravity has
so far been tested only on scales larger than a tenth of a
millimeter \cite{Hoyle01}; possible deviations below that scale can be
envisaged. 

\section{Cosmic Superstrings}

One of the main motivations behind the brane-world scenario was to
try to explain the
vast difference between the Planck scale of gravity of $10^{19}$
GeV and the electroweak scale, which mediates radioactive decay,
of $10^2$ GeV. The idea was to introduce \emph{warped}
space-time. In special relativity we are used to the invariant
distance being given by $ds^2=dt^2-d\boldsymbol{x}^2$. When
space-time is `warped' this becomes
 \begin{equation}
ds^2=e^{-A(\underline{y})}(dt^2-d\boldsymbol{x}^2)
-d\underline{y}^2,
 \end{equation}
where $\underline{y}$ represents the extra dimensions and
$A(\underline{y})$, the `warp' factor, is a known, positive function. The
warp factor is essentially a gravitational red-shift in the
compact directions. In five-dimensional brane worlds this
`warping' of space-time was used to generate a hierarchy of
scales such that  gravity, which propagates on both the brane
and in the bulk, could be at the Planck scale whilst the usual
physics, confined to the brane, could have an energy scale much
less than this. However, the warping  of space-time is more
general than brane-worlds and arises in many string theory
models where there are $6$ compact extra dimensions.  In some
cases the compact dimensions form a simple space such as a
sphere or a torus, with constant warp factor.  But there are
also solutions in which the warp factor varies strongly as a
function of the compact dimensions, $\underline{y}$, with
special regions known as \emph{throats} where it falls sharply
to very small values, as shown in Figure~6. Here the
 \begin{figure}[htb]
 \centerline{\psfig{file=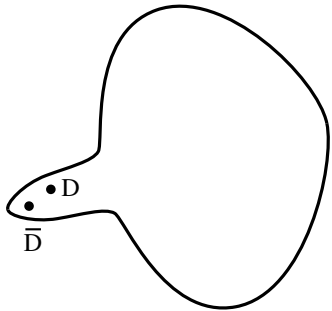}}
 \vspace*{8pt}
 \caption{Space with throat. In the middle we have the compactified
space, with the throat on the left of the figure. The
D$/\overline{\mathrm{D}}$ branes are in the throat.}
 \end{figure}
warp factor is essentially $1$ in the central region and much
less than $1$ in the throat.  This means that, for a
four-dimensional observer the fundamental string mass per unit
length would  appear to be
 \begin{equation}
\mu=e^{-A(y)}\mu_0,
 \end{equation}
where $\mu_0$ is the ten-dimensional scale. Consequently
fundamental  strings may not be at such high energies after
all, even if $\mu_0$ is at the Planck scale. 

Another recent development in string theory is the concept of 
\emph{brane inflation}. The idea of \emph{inflation} is that,
in the very early Universe there was a period of exponential
expansion such that the visible Universe today was
exponentially small at very early times. Since string theory
is meant to be a theory of everything it should also explain
the Universe at very early times, and consequently inflation.
Recently it was discovered that string theory could give rise
to inflation (for a review see \cite{quevedo02}).  Consider
a Universe containing an extra \emph{brane} and
\emph{anti-brane}. The brane and anti-brane are attracted to
each other in the same way as an electron and positron are.
However, when they annihilate they give rise to
lower-dimensional branes \cite{mahbub}  rather than photons.
If the early universe contained an extra brane and
anti-brane separated in the compact dimension, then the
distance between them plays the role of a scalar field,
called the inflaton. The potential energy of these branes
drives the exponential expansion and therefore inflation. As
the branes approach the potential between them becomes
steeper before they annihilate. Once they annihilate lower
dimensional branes are formed. It was shown that D-strings
were generically formed in brane inflation
\cite{sarangitye}.

The most fully developed model of brane inflation involves a
D$3/\overline{\mathrm{D}3}$ at the bottom of a throat,
The D$3$ branes wrap the usual three spatial dimensions, so 
they appear as points in the throat of the extra dimensions,
as shown in Figure~6. The inflation and subsequent brane
annihilation take place in the throat.   After brane inflation
lower dimensional D-branes are formed, also in the throat. In
the above example D1 branes or D-strings are formed.  In
addition fundamental strings, called F-strings can also be
formed. Since the brane annihilation was in the throat where space-time
is highly warped, then the energy scale of these strings is no
longer the Planck scale but at a much lower energy scale.
Estimates give the range to be between
$10^{-11}<G\mu<10^{-6}$ depending on details  of the
theory (see \cite{polchinski04} for a review). 
 
The idea of cosmic superstrings is not new --- they were
proposed as long ago as 1985 \cite{witten} --- but they were
dismissed as at too high an energy scale and unstable.
However, the recent developments in string theory mean that
not merely is this a distinct possibility but also perhaps the
best way of observing string theory. The cosmic superstrings
can have a range of values of string mass per unit length,
$\mu$, compatible with observations
and also the throat essentially provides a stabilizing
potential. These two features circumvent the previous problems
with cosmic superstrings. 

Whilst both D- and F-strings can be formed in brane inflation
they are fundamentally different objects. D-strings are more
similar to the usual cosmic strings discussed in section~2 and
are essentially classical objects, though like all D-branes
they have a conserved charge.  On the other hand F-strings are
quantum mechanical objects.

Finally it now appears that grand unified theories will
almost inevitably give rise to cosmic strings. In the long
road from M-theory to the low-energy physics we observe in the
laboratory, a natural route is  via grand unified theories at
an intermediate stage. In grand unified theories the coupling
constants for the weak, electromagnetic and strong
interactions meet at a high energy scale of about $10^{16}$
GeV. However, the theory is only successful when it is
supersymmetric. A recent study \cite{jeannerot} considered all
possible grand unified theories, up to a certain level of
complexity, with all possible symmetry breaking schemes which
gave rise to the electroweak theory at low energy. The theories 
they considered  all had a period of inflation. The ones that 
were not in conflict with  observations \emph{all} predicted 
the formation of cosmic strings at  the end of inflation. 
Consequently it seems that cosmic strings are  almost inevitable. 

Since the grand unified theories studied were supersymmetric
it seems  natural to study the nature of cosmic strings in
such theories.  Supersymmetric theories give rise to two sorts
of strings, called  D-term or F-term strings \cite{DDT}, where
the D and F refer to the  type of potential required to break
the symmetry, and have nothing directly to do with the distinction
between the D- and F-strings discussed above.
A recent analysis of  supersymmetric theories with a
D-term suggests that D-term cosmic strings may well be
D-strings \cite{dvali}. However, F-term strings  are classical 
objects, and not apparently  related to D- or F-strings.
Nevertheless, the subtle relationships between different string
theories regarded as limits of M-theory may affect some of
these distinctions.

\section{Cosmology of D- and F-strings}

The cosmology of D-strings and F-strings is a little different
from that of ordinary cosmic strings. In section~3 it was
explained that when two cosmic strings meet they intercommute
and that loops are formed by a cosmic string
self-intersecting. For ordinary cosmic strings, the
probability of intercommutation is $1$. This is not the case
for D- and F-strings. For D-strings this is because they can
`miss'  each other in the compact dimension and F-strings are
fundamentally quantum objects so their scattering can only be
computed with a  quantum mechanical calculation. The
probability of intercommuting has  been estimated to be
between $10^{-3}<P<1$ for F-strings and $10^{-1}<P<1$ for
D-strings. Similarly the probability of a string
self-intersecting is reduced. This means that a network of
such strings  could look different from that of cosmic
strings. There are suggestions that such a network would be
denser, with the distance between strings related to $P$, and
slower \cite{dvali&vilenkin, sakellariadou}.  It is likely
that the net result would be  to increase the number of string
loops, despite the reduction in string  self-intersection. A
network of D- or F-strings could also emit  exotic particles
as a result of the underlying superstring theory.  Ordinary
cosmic strings emit particles, but those coming from cosmic
superstrings could have distinctive characteristics. 

Another interesting possibility is that, because they are different strings,
when D- and F-strings meet they are unable to intercommute, instead forming 
a three-string junction, with a composite DF-string, as shown in
Fig.~7. If this were the case then
 \begin{figure}[htb]
 \centerline{\psfig{file=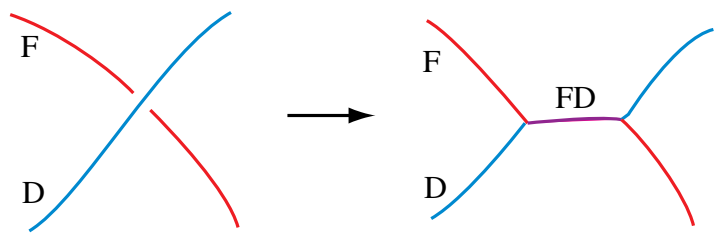}}
 \vspace*{8pt}
 \caption{Crossing of strings of different types.}
 \end{figure}
they would not form loops very effectively. This could be a problem 
since a usual cosmic string network loses energy via loop production.
It is also possible in string theory to have bound states of $p$ 
F-strings and $q$ D-strings! The evolution of such an exotic system 
would be different from that discussed in section~3. It is possible
that a such a network would become \emph{frozen}
and just stretch with the expansion of the universe. Consequently,
there is much to investigate in the cosmology of D- and F-strings.

\section{Observation of cosmic strings}

The most promising way of observing cosmic strings is by
searching for their very characteristic gravitational effects,
either as gravitational lenses or emitters of gravitational
waves.

The gravitational field around a straight, static string is
quite unusual.  Particles in the vicinity feel no gravitational
acceleration, because in general relativity tension is a
negative source of gravity and, since tension equals energy
per unit length, their effects cancel.  Space-time around the
string is locally flat, but not globally flat.  In
cross-section, the space is cone-shaped, with a \emph{deficit
angle} $\delta=8\pi G\mu$, as though a wedge of angle $\delta$
had been removed and the edges stuck together.  The deficit
angle is $\delta=5''\!\!.2 (G\mu/10^{-6})$, so for GUT-scale
strings it is a few seconds of arc.

Thus the string acts as a cylindrical gravitational lens,
creating double images of sources behind the string, with a
typical angular separation of order $\delta$ (see Fig.~8).  A
 \begin{figure}[htb]
 \centerline{\psfig{file=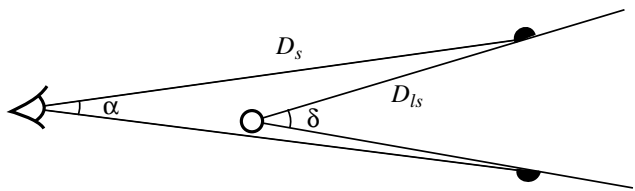}}
 \vspace*{8pt}
 \caption{Gravitational lensing by a cosmic string.}
 \end{figure}
string would yield a very special pattern of lensing.  We
should expect to see an approximately linear array of lensed
pairs, each separated in the transverse direction.  In most
the two images would have essentially the same magnitude.  (The
exception would be if we see only part of one of the images.) 
This is a very unusual signature; most ordinary gravitational
lenses produce images of substantially different magnitude,
usually an odd number of them \cite{gravlens}.

There are several factors that may complicate this picture. 
Cosmic strings are not generally either straight or static. 
Whenever strings exchange partners kinks are created that
straighten out only very slowly, so we expect a lot of
small-scale structure on the strings.  Viewed from a large
scale, the effective tension and energy per unit length will
no longer be equal.  Since the total length of a kinky
string between two points is greater, it will have a larger
effective energy per unit length, $U$, while the effective
tension $T$, the average longitudinal component of the tension
force, is reduced, so $T<\mu<U$.  This means that there is a
non-zero gravitational acceleration towards the string,
proportional to $U-T$.  Moreover, the strings acquire large
velocities, generally a significant fraction of the speed of
light.  If the string is moving with velocity $\boldsymbol{v}$
perpendicular to its direction, the expression for the angular
separation of an image pair is
 \begin{equation}
\alpha = \frac{8\pi GU}{\gamma(1-v_r)}
\frac{D_{ls}}{D_s}\sin\theta,
 \end{equation}
where $D_s$ is the angular-diameter distance of the
source, $D_{ls}$ that of the source from the lensing
string, $\theta$ is the angle between the string and the line
of sight, $\gamma=(1-\boldsymbol{v}^2)^{-1/2}$, and $v_r$ is
the radial component of $\boldsymbol{v}$.

Another very characteristic effect is the distortion of the
CMB produced by a moving string.  A string moving across our
field of vision will cause a blue-shift of the radiation ahead
of it, and a red-shift of that behind, leading to a
discontinuity in temperature of magnitude $\delta T/T=8\pi
GU\gamma v_\perp\sin\theta$, where $v_\perp$ is the component
of the string velocity normal to the plane containing the
string and the line of sight.

Accelerated cosmic strings are sources of gravitational
radiation.  The most important signal would come from special
places where the strings are moving exceptionally fast. 
Loops of string undergo periodic oscillations, with a period
related to the size of the loop.  The dynamical equations
predict that during each oscillation there will be a few
points at which the string instantaneously forms a cusp, where
it doubles back on itself (see Fig.~9).  In the neighbourhood
 \begin{figure}[htb]
 \centerline{\psfig{file=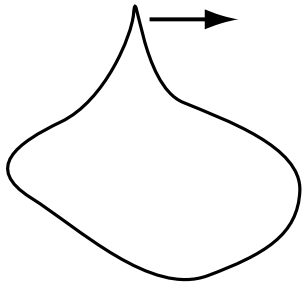}}
 \vspace*{8pt}
 \caption{Cosmic string loop with a cusp.}
 \end{figure}
of the cusp, the string velocity approaches the speed of
light.  Such an event would generate an intense pulse of
gravitational radiation, strongly beamed in the direction of
motion of the cusp.  If massive cosmic strings do indeed
exist, these pulses are likely to be among the most prominent
signals seen by the gravitational-wave detectors now in
operation or planned, in particular LIGO and LISA.

This effect has already provided a stringent, though indirect,
limit on the value of $G\mu$.  This comes from observations of
the timing of millisecond pulsars.  Gravitational waves
between us and a pulsar would distort the intervening
space-time, and so cause random fluctuations in the pulsar
timing.  The fact that pulsar timings are extremely regular
places an upper limit on the energy density in gravitational
waves, and hence on $G\mu$.  The upper limit is of order
$10^{-7}$, though there is considerable uncertainty because
this depends on assumptions about the evolution of small-scale
structure. 

For cosmic superstrings the situation is similar. However,
because the intercommutation probability is less than unity
the evolution of the network is a little different, resulting
in a denser, slower network of strings with more cusps on it.
A recent study suggests this could enhance the gravitational
radiation emitted  \cite{damour&vilenkin04}
and that such strings could be  detectable with the
gravitational-wave detectors in the near future.  Seeing such
cosmic strings could provide a window into superstring theory!

\section{Recent observations}

\subsection{A possible example of lensing by a cosmic string}

One exciting recent piece of evidence was the observation of a
possible example of cosmic-string lensing by a
Russian--Italian collaboration, between the Observatory of
Capodimonte in Naples and the Sternberg Astronomical Institute
in Moscow.

What Sazhin \emph{et al.}\ \cite{Saz+03} saw was a pair of
images of apparently very similar elliptical galaxies, both
with a red-shift of 0.46, separated by $2''$.  The images have
the same magnitude and the same colour --- the magnitudes in
three separate wavelength bands are equal within the errors. 
They could of course be images of two distinct, but very
similar galaxies that just happen to lie very close together. 
Close pairs are not unusual, but it would be a remarkable
coincidence to find two so similar so close together.  The
images could also be due to lensing by some more conventional
foreground object, but the authors show that it would have to
be a giant galaxy, of which there is no sign.

They conclude that the most likely explanation is lensing by a
cosmic string.  If so, the observed separation requires that
$G\mu>4\times 10^{-7}$, which is at least marginally in
conflict with the upper limits from CMB anisotropy and pulsar
timing observations.  However, it should be remembered that
this is actually a limit on the effective quantity $GU$ rather
than $G\mu$ so there is at least some room for manoeuvre.

Another important piece of evidence comes from a later study
by the same authors of the surrounding region \cite{Saz+04}. 
If the image pair is due to lensing by a cosmic string, one
would expect other lensed pairs in the vicinity, along the
line of the cosmic string.  The authors searched for such
pairs in a $16'\times16'$ patch of sky around the original
image pair (which is called CSL-1, the
\emph{Capodimonte--Sternberg Lens Candidate no.~1} --- there
are three others yet to be analyzed).  Among the roughly 2200
galaxies within this patch, they found 11 very likely
candidates for lensed pairs, based on separation and colour
matching.  They estimate that a string should produce
somewhere between 9 and 200 lensed pairs (depending on its
configuration), while they should expect no more than 2 due to
conventional lensing.  So this adds weight to their
interpretation, though they emphasize that the identification
needs to be confirmed by a spectroscopic analysis of the pairs.

We can learn a lot from the distribution and alignment of the
candidate pairs.  A straight string should produce a linear
array of lensed pairs, with the pairs separated in the
transverse direction.  A picture of the six brightest pairs
\cite{Saz+04b} (see Fig.~9) does not show such a sharp
 \begin{figure}[htb]
 \centerline{\psfig{file=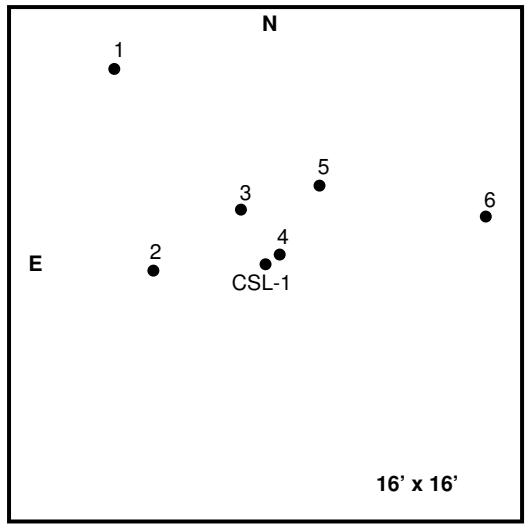}}
 \vspace*{8pt}
 \caption{Positions of candidate lensed pairs in the vicinity
of CSL-1 \cite{Saz+04b}.}
 \end{figure}
concentration, but nor do they seem to be randomly scattered. 
The position angles of the pairs nos.\ 2,3,5,6 do suggest that
they could line up on a smooth curve of string \cite{Saz04}. 
The others could perhaps be fitted to a string with a couple of
kinks.  The important test of this idea will come from a
spectroscopic analysis of the candidate pairs, to show whether
they are indeed images of the same object.

\subsection{Possible lensing by an oscillating loop}

The other intriguing development is an analysis by Schild
\emph{et al}.\ \cite{Sch+04} of brightness fluctuations in a
very well known gravitational lens system that has been studied
extensively for 25 years.  The system is famous because it has
been used to provide an estimate of the Hubble constant,
independent of other distance measurements.  It consists of a
pair of images, which are known to be images of a single
quasar, because of the constant observed time delay:
brightness fluctuations in image $A$ are generally followed by
similar fluctuations in
$B$ 417.1 days later.  The lensing in this case is due to a
clearly visible foreground galaxy, and the time delay occurs
because one light path from the quasar is a little more than
one light year longer than the other.

In addition to the correlated fluctuations there are
independent fluctuations of the two images, primarily caused
by microlensing by individual stars in the foreground galaxy,
that is, lensing in which different images are not resolved. 
What Schild \emph{et al}.\ have have found in data from
observations in 1994--95 is an apparent sequence of synchronous
fluctuations in both images with an amplitude of about 4\% and
no time delay.  They see a sequence of three or four
oscillations with a period of about 80 days.

If these oscillations do have a common origin, they must be
due to some object quite close to us.  One possibility would
be lensing by a binary star, but to get the right period and
amplitude the stars would need be among our near neighbours
and to have masses of around eighty solar masses.  It is
inconceivable that such massive stars so near us could have
escaped detection.

Another possibility is an oscillating loop of cosmic string. 
Since the period is proportional to the length of the string
loop, the required 80-day period would imply a length of 160
light-days.  The loop would probably be moving with a
substantial fraction of the speed of light, so it would only
remain between us and the source for a year or so; thus it is
not surprising that only a few oscillation cycles were
seen.  The apparent angular size would need to be somewhat
less than the separation between the images (otherwise there
would be sharp spikes of intensity when the string actually
crossed one of the paths).  On the other hand the loop cannot
be very far away, otherwise the required value of $G\mu$ would
be impossibly large.  In fact, it must be well inside our
galaxy.  Since we know from CMB and gravity-wave limits that
loops of this size must be quite rare, to find one so near
would be remarkably fortuitous.

Of course, the sequence of synchronous fluctuations might just
be a coincidence, but the authors argue that the probability
of seeing three or more synchronous oscillations by chance is
quite low.  Nevertheless, this may be the simplest
explanation; we need a proper statistical analysis of how
likely it is that this should have happened by chance at some
time during the many years of observation.

\section{Conclusions}

As we have described, recent developments in fundamental
string theory, especially the brane-world concept, have
greatly extended the range of different kinds of cosmic
strings that might have been formed in the very early
history of the universe.  There have been intriguing
hints of observations that might be signatures of cosmic
strings.  Further work in the near future should clarify
their status.  Even if these observations turn out to have
more prosaic explanations, the quest for evidence of cosmic
strings will certainly continue, in particular via searches
for the gravitational waves they emit.  The very
characteristic signature of emission from a cusp
\emph{might} be detectable even by the present generation
of gravitational-wave detectors, and certainly by the
next.  This may well provide the first direct evidence for
an underlying superstring or M-theory.

\section*{Acknowledgments}

We are indebted for valuable comments and discussion to Ana
Ach\'ucarro, Levon Pogosian, Fernando Quevedo, Mairi
Sakellariadou and Miguel Sazhin.  The work reported here was
supported in part by PPARC, and in part by the ESF through the
COSLAB Programme.

\end{document}